\documentclass[journal]{IEEEtran}
\usepackage{amsfonts}
\usepackage{verbatim}
\usepackage{mathrsfs}
\usepackage{amssymb,amsmath}
\usepackage[noblocks]{authblk}
\usepackage{stfloats}
\usepackage{algorithm}
\usepackage{algorithmic}
\usepackage{cite,url,subfigure,graphicx,wrapfig,verbatim,amsmath,amsfonts,amssymb,color, epsfig,graphics,caption}
\usepackage{theorem}
\usepackage{array,color}
\usepackage{multirow}

\theoremheaderfont{\normalfont\bfseries}

\newcommand{\eg}{e.g.}
\newcommand{\ie}{i.e.}
\newcommand{\etc}{etc}
\newcommand{\etal}{\textit{et al. }}

\begin{document}
%TC:ignore
\title{Actions at the Edge: Jointly Optimizing the Resources in Multi-access Edge Computing}
\author{\IEEEauthorblockN{Yiqin Deng, Xianhao Chen, Guangyu Zhu, Yuguang Fang,~\IEEEmembership{Fellow,~IEEE}, Zhigang Chen, Xiaoheng Deng
\thanks{Corresponding author: Xianhao Chen.}
\thanks{Yiqin Deng, Zhigang Chen and Xiaoheng Deng are with School of Computer Science and Engineering, Central South University, Changsha, China (email: \{dengyiqin, czg, dxh\}@csu.edu.cn). }
\thanks{Xianhao Chen, Guangyu Zhu and Yuguang Fang are with the Department of Electrical and Computer Engineering, University of Florida, Gainesville, FL, USA (email: \{xianhaochen@,gzhu@, fang@ece.\}ufl.edu).}
%\thanks{Haichuan Ding is with University of Michigan, Ann Arbor, MI, USA (email: dhcbit@gmail.com).}
%\thanks{Pan Li is with the Department of Electrical Engineering and Computer Science, Case Western Reserve University, Cleveland, OH 44106, USA (email: lipan@case.edu).}
}}

\markboth{Multi-access Edge Computing}{Actions at the Edge: Jointly Optimizing the Resources in Multi-access Edge Computing (MEC)}

\maketitle

\begin{abstract}
Multi-access edge computing (MEC) is an emerging paradigm that pushes resources for sensing, communications, computing, storage and intelligence (SCCSI) to the premises closer to the end users, \ie, the edge, so that they could leverage the nearby rich resources to improve their quality of experience (QoE). Due to the growing emerging applications targeting at intelligentizing life-sustaining cyber-physical systems, this paradigm has become a hot research topic, particularly when MEC is utilized to provide edge intelligence and real-time processing and control. This article is to elaborate the research issues along this line, including basic concepts and performance metrics, killer applications, architectural design, modeling approaches and solutions, and future research directions. It is hoped that this article provides a quick introduction to this fruitful research area particularly for beginning researchers.

\end{abstract}

%TC:endignore

\section{Introduction}
\begin{figure}
\centering
\includegraphics[width=0.5\textwidth]{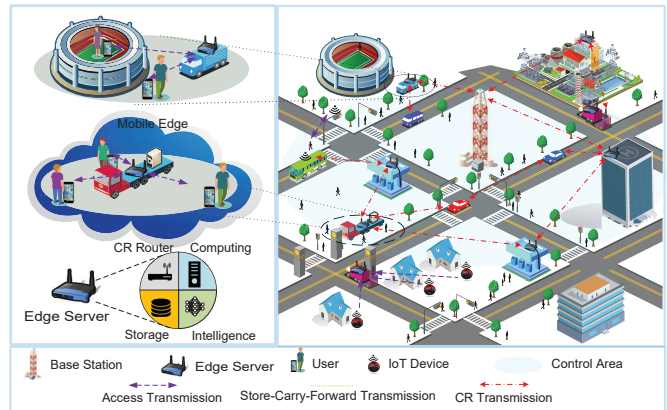}
    \caption{A typical scenario for  MEC.}\label{fig:mec_app}
\vspace*{-0.1in}
\end{figure}

Multi-access edge computing (MEC) is an emerging service network architecture to push powerful services to the proximity of end users, which includes not only the network services as proposed in the formerly mobile cloud computing, but also sensing, computing, storage and intelligence. Thus, service requests from customers could be flexibly provided at edge nodes, e.g., base stations (BSs), access points (APs), and roadside units (RSUs). Imagine that in a smart city envisioned in the last few years, if customized devices with powerful sensing, communications, computing, storage, and intelligence (SCCSI) capability are installed in or co-located with existing infrastructure (\eg, BSs, APs, RSUs, rooftops, lamp posts, \etc) or vehicles (\eg, public transits, vehicles, \etc), then we will have a powerful MEC system, consisting of mobile infrastructure to take advantage of both spectrum and mobility opportunity, to provide needed SCCSI services \cite{ding2017cognitive2}. Due to the omnipresence of vehicles (public or private) in a city, their mobility could easily reach the proximity of end users, \ie, the edge. Moreover, by shifting processing/computing power from the remote cloud to locally situated, possibly reconfigurable, edge servers (ESs), MEC helps significantly mitigate traffic congestion over backbone networks and reduce end-to-end (e2e) latency for many emerging capability demanding applications, enhancing QoE for end users. This vision can be easily illustrated in Figure \ref{fig:mec_app} where powerful ESs are populated in a smart city to build a robust service network for SCCSI services.

Due to its tremendous potential to revolutionize the telecommunications and computing industries, there is a surge in MEC, particularly in wireless industries \cite{koketsurodrigues2021offloading}. The early concept of MEC was conceived at the early stage of wireless industries, particularly when mobile devices had limitations on resources on their own and there was a need to pull resources from mobile devices in close proximity together for sharing, leading to the cloudlet or the later mobile edge computing (also termed MEC). Following the developments of wireless mesh networks and cloud computing, where specialized powerful routers or computing servers were deployed at strategic locations to beef up the communications and computing at the edge, i.e., push communications and computing services closer to the end users, the true marriage of communications and computing started to emerge, leading to the birth of powerful MEC technologies. In light of integration of AI into future wireless technologies and real-time intelligent control, 5G systems and beyond (simply 5G+) have already taken edge computing capability into their design consideration, positioning it as the enabler to intelligentize the vertical industries.

Although there are extensive works on MEC done in the past few years, many open problems are still unresolved. The challenges lie in the fact that task offloading involves many design factors, not only computing resources, but also transmission capability. A task to be computed has to be successfully delivered to an ES and then computed within the timeline required by the specific applications. Besides, both transmissions and computing systems should have sufficient storage to buffer the task for resource scheduling. Thus, an effective MEC system has to optimally coordinate the distributed spectrum resource, computing resource, and storage resource under the constraints on power consumption and latency.

In this paper, we plan to articulate the problems and challenges in MEC, discuss the possible models and solution approaches, and then identify the future research directions. Different from some existing surveys on MEC which attempt to provide comprehensive summary of research efforts in this area~\cite{mao2017survey,abbas2017mobile, mach2017mobile}, this paper targets at design taxonomy, research problems, typical architectures, and potential solutions by jointly considering communication and computing resources from an end-to-end perspective. Moreover, we will offer enlightening thoughts on several key issues, including the use of queuing model and multi-hop offloading for MEC scenarios.

\begin{comment}
The remainder of this paper is organized as follows. In the next section, we present a few killer applications of MEC to further motivate the research. In Section III, we offer a general framework for MEC, fundamental problems and performance metrics. In Section IV, we discuss the key design issues based on a classification of system settings and modeling approaches, followed by a case study in Section V. In Section VI, we point out a few important research directions in MEC and conclude the paper in the last section.

\end{comment}

\section{Selected Killer Applications}

To demonstrate how important MEC is, in this section, we present an non-exhaustive list of important killer applications to demonstrate how MEC can be leveraged to boost both communications and computing performance for better users' QoE.

\vspace{-0.15in}
\subsection{Connected and Autonomous Driving}
%Although CAD heavily rely on self-equipped sensing and computing capability, many assistive supporting information systems must be in place in order to maintain the needed safety requirement. For example,
%, particularly for VR/AR (virtual reality/augmented reality) navigatio
The one on top of attractive applications for MEC should be connected and autonomous driving (CAD). CAD does need near real-time map for collision avoidance and pedestrian safety, thus the simultaneous localization and mapping (SLAM) is crucially important. Yet, SLAM demands timely sensing (\eg, video) and fast computing capability to perform fine-grained video analytics, which can hardly be done by connected and autonomous vehicles (CAVs) on time. Besides, any highly complex computing and processing shifted from CAVs to the edge could significantly reduce the cost of CAVs. One may argue that such tasks can be offloaded to the cloud, but yet the latency may pose serious problem for CAD. Thus, the only alternative is an effective MEC.

\vspace{-0.15in}
\subsection{Industrial Internet of Things}

Many real-time applications such as industrial Internet of Things (IoT) do require low latency for real-time scheduling and control. For example, future smart manufacturing may demand timely sensing to collect much needed data, deliver them to a certain computing facility to carry out necessary computing for intelligence extraction, and then take timely control actions. %In fact, many industrial control systems do require timely signal processing for intelligent control.
Due to the tight latency requirement, MEC seems to be the only choice.

\vspace{-0.15in}
\subsection{Video Surveillance}

To build a smart city, public safety is one of the important design goals. To this end, video surveillance cameras may be deployed. Due to large volume of video data, it is impractical to upload all the raw data to the cloud to carry out video analysis. More often timely video analytics may have to be done within certain time frame in order to fight possible crimes in tough neighborhoods. This is exactly what MEC could come to rescue. %which should be investigated carefully.
%Besides, intensive computing tasks may be offloaded to the places where computing capability and storage are sufficient or devices that can help conduct video analytics via crowdsourcing

\vspace{-0.15in}
\subsection{Smart and Connected Health}

Information and communications technologies (ICT) and artificial intelligence (AI) have revolutionized what we care for people's health, and smart and connected health (SCH) is a great effort towards this goal. For people who have chronic or life critical diseases, SCH could help collect vital signals under 24/7 clock, perform timely signal analysis to monitor potential changes in signal pattern by applying sophisticated AI algorithms such as machine learning (ML), deep learning (DL), and federated learning (FL), \etc. Due to this periodic signal collection, huge volume of data will be generated, resulting in big data to be handled, and it may not be necessary to always interact with the cloud, but yet timely data analytics may have to be done locally. Thus, MEC provides another rich application to address important problems.
%It can also be applied to collect the vital life data for elderly people on constant basis to perform signal change detection for smart aging or even early Alzheimer detection.

\section{General MEC Framework}

With such diverse applications of MEC, the problem is how to design such an MEC system to suit a practical application. In this section, we  first present a general architectural design guideline for MEC, and then identify the fundamental design issues under the MEC framework to improve users' QoE. Afterwards we discuss three critical metrics to measure the performance of MEC. %for specific problems.

%\begin{wrapfigure}{r}{0.3\textwidth}
\begin{figure}
\centering
\includegraphics[width=0.5\textwidth]{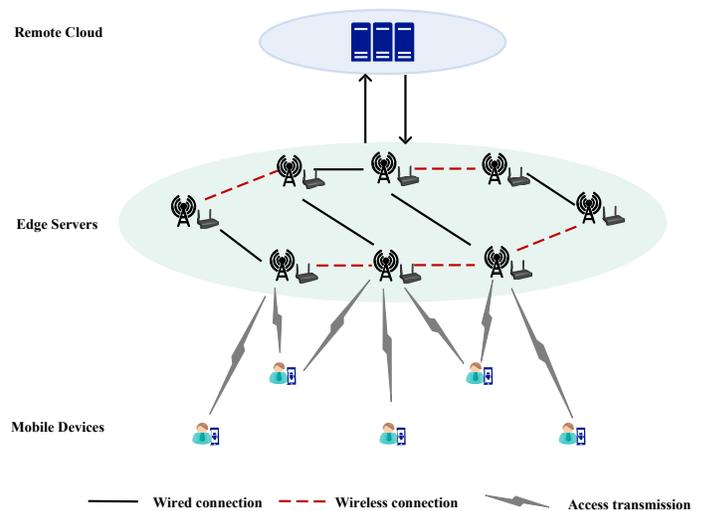}
\centering
    \caption{The generic architecture of an  MEC.}\label{fig:mec}
\vspace*{-0.1in}
\end{figure}
%\end{wrapfigure}

\vspace{-0.15in}
\subsection{Architectural Design for MEC}

An MEC consists of the cloud, edge servers (ESs) and mobile devices (MDs) as shown in Figure \ref{fig:mec}. A set of ESs are populated in the area of service (AoS) while MDs roaming in the AoS could offload their tasks to ESs for computing (for signal processing or machine learning). An ES is typically customized with powerful communications capability (\eg, cognitive radios), high computing power and sufficient storage space, running various AI algorithms, which is typically termed {\em CCSI} capability. If it is also equipped with internal or external sensing features or interconnected with sensors in its proximity, we could enable MEC the SCCSI capability. An MD could access the computing, storage and intelligence resource at its proximity if reliable communication link(s) can be established to an ES (or a subset of ESs). ESs can be connected via high-speed links (solid lines) and/or  wireless links (dashed line), depending on the deployment strategy for ESs.

\vspace{-0.15in}
\subsection{Fundamental Problems}
The fundamental problem in MEC is how to facilitate MDs to access the services at ESs with desired quality of service (QoS). This mainly includes the following few design issues.

\textbf{Computation Offloading}. An MD could offload computing task either wholly or partially, depending on both local computing resource in the MD and the channel condition from the MD to an ES(s) it could reach. However, this problem is highly challenging since both the channel condition and computing resource (at both the MD and the chosen ES(s)) should be taken into consideration. Moreover, partial offloading decisions involve additional constraints, \eg,  whether the tasks are allowed to be split into variable ratio or not, which is largely ignored in most existing works.

\textbf{Resource Allocation}. After task offloading decision is made, the next design task is to allocate enough communication resource between an MD and selected ESs and sufficient computing power at both the MD and selected ES(s). An efficient resource allocation has to be designed due to the limitation of resources in MEC.

%\textbf{Caching and Prefetching}. Although storage can be regarded as a type of resource at ESs, due to its importance, we deal with it separately when caching and prefetching is considered. With MEC is in place, caching can be handled effectively via appropriate content distribution, so that the cached data items can be easily accessed by end users. On the other hand, computing typically relies on the placement of non-trivial amount of data, such as database and models. Thus, how to design effective caching or prefetching strategies is challenging, which have become important design tasks for 5G+ lately.

\textbf{Caching and Prefetching}. Although storage can be regarded as a type of resource at ESs, due to its importance, we deal with it separately from resource allocation when caching and prefetching are considered. Caching/prefetching at the edge (or edge caching) is different from the traditional caching in that the goal here is to enable corresponding computing tasks to be executed efficiently at ESs and is often jointly designed with task offloading to maximize MEC performance. Moreover, computing at the edge typically relies on the placement of non-trivial amount of data for training and ML models for edge intelligence while traditional caching works mostly for content distribution. Thus, how to design effective caching or prefetching strategies in MEC system is significantly different from traditional approach and highly challenging.

\textbf{Energy Conservation}. Since MDs tend to be battery powered, energy conservation is always an important design consideration. The energy consumption comes from both transmissions and computing. A computing task can be done locally at an MD or offloaded to an ES or jointly done at both sides in order to conserve energy on the MD.

\textbf{Mobility Management}. Many network entities such as MDs or mobile ESs may be subject to mobility, which will result in degraded communications quality or longer e2e latency. How to manage the mobility to sustain continuous communications services or effective computing is an important design issue. To keep the task computing ESs closer to the end users on the move, mobility management (\eg, user tracking) and task computing migration (\eg, virtual machine migration) must be taken into consideration in practical applications \cite{yuan2020joint}.

\vspace{-0.15in}
\subsection{Performance Metrics}
To assess the performance of MEC, we typically need some key performance indices (KPIs). In this subsection, we present three essential performance metrics for the aforementioned problems from the customers' perspective (\ie, latency and energy consumption at MDs) and service providers' perspective (\eg, task completion rate or simply throughput), respectively.

\textbf{Latency}. MEC targets at providing timely access to edge services, and hence latency is definitely the first metric to guarantee. Latency here means that the time between the instant that an MD submits its computing task to the instant it obtains the computing result of the task. Since this definition is from the perspective of an MD, it is indeed the end-to-end (e2e) latency. Since the MD may make a decision based on the result of the completed task within certain time limit, latency is indeed an important metric.

\textbf{Energy Consumption}. Energy consumption either due to task transmissions or due to task computation locally at an MD is always a major concern for resource-constrained systems. Since many end users tend to use their smart devices for the applications of their interests, it is always good idea to save the battery life of their MDs. Thus, the total energy consumption at an MD for a task completion under a task offloading scheme is another metric to be considered.

\textbf{Throughput (task completion ratio)}. To maximize the revenue, a service provider for an MEC system tends to utilize system resources as efficiently as possible to accommodate customers to maximize the completed tasks, which indicates that the system throughput (or task completion ratio) is a critical metric for MEC. To boost the throughput, load balancing across multiple ESs is an effective way since it improves service availability by spreading the workload based on load status in terms of both communication and computing. To optimize the aggregated throughput and users' QoE, MD's tasks have to be transferred to the edge server with lower computing load via less congested links.
%Specifically, whenever the associated edge server for a user (MD) is overloaded either due to the network congestion or excessive computing, the task transmission time and/or task computing time could be prolonged, violating the latency constraint.
%Thus, the throughput is indeed another important metric to assess the MEC design.

\section{Resource Allocation and Load Balancing: An End-to-End Perspective}

With all basic concepts clarified, based on the current research activities, we are now ready to categorize major research problems on MEC. In this paper, we focus on two major research tasks in MEC, namely, resource allocation and load balancing. Resource allocation tends to address how to allocate resources for end users from one service facility (e.g., an ES) while load balancing focuses on how to balance the workloads over multiple service facilities (e.g., ESs).  Typically, depending on specific applications, resource can be different. Although our envisioned MEC consists of SCCSI services, in this paper, we focus mainly on communications and computing. We will articulate the research problems involving with spectrum allocation, computing offloading, and the related load balancing under different system setup settings.
%Since MEC systems consist of end users who request for services and edge servers/cloud which provide services, we will elaborate the research activities according to the system setup setting.

\vspace{-0.15in}
\subsection{Single-Edge and Single-MD Case (SESM)}

We start with the simplest case first to articulate the problem. For an MEC system with SESM, the problem is whether to offload computing from the mobile user device (MD) partially or wholly to the edge server (simply ES). The decision depends on various factors of concern. If we are concerned with the MD's energy conservation, we may compare the energy consumption on data offloading (depending on spectrum availability and channel conditions between MD and ES) and the computing power consumption at MD if computed locally. This problem can also be formulated as partial computing offloading problem by dividing the whole computing task into two parts to be performed at MD and ES separately if a task is divisible. If a computing task is latency sensitive, then the problem can be formulated as MD energy minimization under e2e latency constraint. In \cite{han2020joint2}, we formulated a joint optimization problem to minimize long-term power consumption at the MD under resource constraints on communications and computing and applied Lyapunov optimization to obtain approximate solutions.
%In this case, the offloading can be formulated as an MD energy minimization problem under some constraints.

% summary of literature: may just need a couple of representative works.

If latency is not a deciding factor, \eg, for delay-tolerant computing, a computing task can be queued either at MD waiting for enough spectrum and/or better channel condition for opportunistic offloading or queued at an ES for sufficient computing resource to perform opportunistic computing.

\vspace{-0.15in}
\subsection{Single-Edge and Multi-MD Case (SEMM)}

This is probably the most common scenario investigated in the current literature. Under this scenario, we have only one single server to provide computing services and have a single shared spectrum to provide offloading for all MDs. If too many users attempt to offload, it may not be worthwhile for an MD to offload its task to the ES due to either spectrum shortage or computing overload at the ES, because this may cause excessive energy consumption when uploading the task or intolerable latency due to communication/computing queueing delay. Thus, the problem has become more complicated. Due to the competing nature for both spectrum and computing, energy minimization and latency minimization for task computing should be formulated carefully. The coupling of task offloading and computing may impact the final decision on whether task computing is carried out either locally or at ES.
%Therefore, an appropriate optimization may have to be formulated and investigated carefully.

% Summarizing the papers on this topic: seem as before.

If queueing is allowed, various queueing models can be developed for conserving energy on MDs or reducing latency for task computing. One MD may choose to opportunistically offload tasks only when channel to the ES is good or choose to offload only when the computing work load at ES is light.

\vspace{-0.15in}
\subsection{Multi-Edge and Single-User Case}

This scenario is not really practical, but may be useful to capture certain essential features of practical problems. It may be used to model the decision process when an MD decide where to offload its computing task, given the MD could offload its task to a subset of ESs. The set of problems for this case can be classified into the following cases. i) Given the channel conditions and computing load on each ES, minimize the MD's energy consumption under latency constraint; ii) Given the channel conditions, computing load on each ES, and residual energy in MD, split computing tasks into local computing and edge computing to maximize the task throughput under latency constraint and MD energy constraint; and iii) Given the channel conditions and computing load on each ES, maximize the task throughput under latency constraint. With multiple edge servers available, the computing tasks from the MD have to consider the load balancing in terms of both communications and computing. In \cite{han2019offloading}, we investigated this problem and identified the bottleneck in the offloading process.

% % Review literature if any.

% When queueing is allowed, we can investigate all the aforementioned optimization problems by leveraging the opportunistic communications and computing, which will boost performance further.

\vspace{-0.15in}
\subsection{Multi-Edge and Multi-User Case}

This is probably the most general scenario that we typically encounter in MEC systems. In the nutshell, the problem at hand is to match the computing tasks with appropriate ESs to optimize the desired design objective, given the channel conditions between MDs and ESs and the computing work loads on ESs. The design objective can be either minimizing MD energy consumption or maximizing the task throughput, both under task latency constraint. Some researchers have also proposed to minimize latency under local power constraints. However, it is noted that minimizing latency may not generate enough incentive for MEC operators as end users will be satisfied and pay the same bills as long as the latency requirement is met. Many papers addressed simplified version of the aforementioned problem: given a set of computing tasks to be offloaded, match the ESs under the latency constraints, which has not taken the task arrival uncertainty into consideration, and hence is not practical at all. How to formulate the right set of optimization problems is of paramount importance, yet highly challenging, particularly under stochasticity in task arrivals, spectrum availability, and computing workload.

Besides, most papers focus on offloading via one-hop transmissions between MDs and ESs. Recently, we have started investigating whether we could leverage multi-hop transmissions to balance the computing workload at ESs and transmission workload from MDs to ESs while meeting the latency constraints. In this subsection, we elaborate research activities based on whether offloading is done via single-hop or multi-hop.

\subsubsection{Single-hop offloading}

Most current research activities focus on single-hop offloading in the sense that an end user is seeking a powerful edge in its proximity and offload its computing tasks directly \cite{poularakis2020service}.

Some works assume that computing resources at ESs are sufficient and formulate the optimization problem either to maximize the sum throughput or to minimize the energy consumption at MDs. Some other works assume that communication resources are sufficient and formulate the optimization problem to balance the computing workload at ESs. Both scenarios, although reducing the problems significantly, are not really practical.

In \cite{poularakis2020service}, Poularakis \etal studied probably the most general problem to minimize the workload offloaded to the cloud under constraints on communications, computing and storage at ESs, but did not taking the queueing model into consideration. Recently, we formulate the optimization problem to maximize the task completion rate (throughput) by balancing the workloads for both communications and computing under e2e latency constraints \cite{deng2021throughput2}. The intuitive idea is motivated by the observation that when we offload a task for computation, we should not only offload it to an ES that does have sufficient computing resource to complete the task, but more importantly should have enough communications resource to upload it in the first place in order to complete it. Thus, in \cite{deng2021throughput2}, under certain assumption on the stochasticity of involved processes (task arrivals, random channels and varying computing power), we use the tandem queue to model the joint system of communications and computing, and find the (approximate) solution, which represents significant departure from the main stream research.

\subsubsection{Multi-hop offloading}

Although it is much easier to formulate the optimization problem under one-hop offloading, in reality, due to the locality of end users and their running applications, task offloading may be difficult to achieve due to heterogeneity in the availability in spectrum and computing resource at different location and at different time. Some ESs may have abundant computing resource, but do not have enough spectrum resource at their proximity to receive tasks, while some of them do have sufficient spectrum resource to receive tasks, but do not have enough computing power. Yet, task completion does demand both sufficient spectrum resource and computing power. Thus, if there are ways such as emerging omnipresent vehicular communications systems to relay tasks from congested location to less congested area via multi-hop relays, it could potentially boost the task completion rate (throughput). For example, in~\cite{deng2021leverage}, we have proposed to leverage vehicles to transport tasks from congested areas to less congested areas to significantly increase the task throughput. Data transportation can of course be carried out either through store-carry-forward mechanisms or via V2V relays over (opportunistic) vehicular communications, which leverages both the mobility opportunity and the spectrum opportunity.

One special scenario is that all ESs in a specific area are interconnected via high-speed reliable links like cable or optical fibers so that the workload in the network of ESs can be balanced directly among them. In this case, the aforementioned task offloading problem can be boiled down to the load balance problem over spectrum resource via relaying nodes, that is, tasks can be uploaded to any ES. Thus, the problem to be solved is to design anycast routing  schemes over potentially wireless ad hoc networks with latency constraints.

To conserve MDs' energy under multi-hop task offloading strategies, end users could search for agents, \eg, nearby powerful relaying nodes, to help relay their tasks to be computed at appropriate ESs. This would lead to another rich set of optimization problems for MEC task offloading.

\subsubsection{Queueing Consideration}
It is typically observed that many prior works have not really taken specific queuing models into the optimization to leverage the opportunistic communications and computing~\cite{poularakis2020service}. We recently utilize the tandem queue models to capture the stochastic nature of task arrivals and service (transmission time and computing time)~\cite{deng2021throughput2,deng2021leverage}. Queuing scheduling is unavoidable in any service systems which can be utilized to increase the resource efficiency and enhance QoE. This is particularly true for joint distributed communications and computing systems for task offloading which can be better optimized by leveraging both mobility opportunity and spectrum opportunity as in vehicular MEC. Unfortunately, the complexity becomes overwhelming when queuing models are taken into consideration for task offloading in MEC, which demands further investigation.

\section{Future Research Directions}

Although we have already mentioned some research challenges and possible solutions under various MEC system settings, there still exist many challenging research problems in this area. Here we only list some important research directions we have identified that are urgently needed to be resolved.

\vspace{-0.15in}
\subsection{Place Edge Servers for Effective MEC Design}

Depending on the cost in response to the coverage requirement, ESs have to be appropriately placed in the service area of interest. The closely related research is on placement of ESs aiming at minimizing the access latency from MDs to ESs or energy consumption, where communication-related service quality metrics have been used to address the coverage issue~\cite{zhang2021joint}. However, in terms of computing, how to quantify the service coverage from an end-to-end perspective is still unclear. One possible solution is to consider some use cases such as smart city applications, and formulate the optimization problem to minimize the computing service outage (\ie, the probability that a computing task cannot be completed within the latency requirement).
%We may also learn from the BS placement from cellular systems, come up with heuristics to ensure the coverage probability for computing at each ES, and then extend the result to investigate the global coverage in the service area.

\vspace{-0.15in}
\subsection{Take Queuing Models into Account}

To simplify the modeling process or reduce the complexity in optimization in MEC systems, queue models for task uploading and computing are typically ignored and even if they are considered, they are just simplified versions, which may not be practical. In reality, the performance highly depends on how a task transmission is done and/or how its computing is scheduled, which is particularly true when leveraging geographical differences of users and ESs as in multi-hop offloading scenarios \cite{deng2021leverage}. Therefore, how to incorporate the appropriate queuing models into optimization is an important yet challenging research task.

\vspace{-0.15in}
\subsection{Leverage In-Network Resources}

Current wireless networking environments are now equipped with rich powerful SCCSI capability which could be leveraged to fulfill needed service provisioning. For example, many today's vehicles, particularly the emerging connected and autonomous vehicles (CAVs), are installed with powerful computing and storage units and sensors, but they are not intended for communications and computing services for external applications. Yet, their omnipresence in smart cities could definitely help provide additional MEC services \cite{ding2019beef2}. How to take advantage of such in-network resources will form a fruitful future research direction.

\vspace{-0.15in}
\subsection{Take Advantage of Mobility and Spectrum Opportunities}

Many powerful communications and computing devices with more rich storage tend to be mobile due to shared mobility, user mobility, or things' mobility in general. For example, public transits carry users with SCCSI capable devices or they themselves are equipped with powerful SCCSI capability, and hence their routine or planned routes would enable them to regularly carry such SCCSI services, a rich set of MEC services that can be leveraged. Moreover, wireless spectrum is location-based, and their utility is shown to be opportunistic spatially and temporally \cite{ding2017cognitive2}. Thus, for an effective MEC system to be designed with specific missions, it is possible to leverage both spectrum and mobility opportunities to boost its service performance without adding extra infrastructure and/or resource. If user owned equipments are used, crowdsourcing and the corresponding incentive mechanism design can be incorporated into this design.

\vspace{-0.15in}
\subsection{Utilize Machine Learning to Improve MEC}

Machine learning (ML) has emerged as  the panacea to solve many difficult engineering problems which cannot be resolved easily with traditional mathematical tools. Since computing at edge via an MEC system involves with factors such as computing load, transmission environments and network load, user device capability, and complicated queue scheduling for both communications and computing, traditional mathematical modeling based on first-principles will result in highly complicated optimization problem with high dimension (\ie, the curse of dimensionality). ML may offer certain level of model reduction by capturing only the important features according to the specificity of the involved applications \cite{rodrigues2019machine}. Moreover, for cases that traditional modeling is hard to handle, model-free ML may be evoked to find an approximate model (data-driven approach). For example, the multi-hop offloading problem may be solved via reinforcement learning, which is currently under  investigation.

\vspace{-0.15in}
\subsection{Utilize MEC to Improve Machine Learning}

%ML typically consumes intensive computing resource, particularly for real-time or delay-sensitive tasks as in connected and autonomous driving, which is why MEC was introduced \cite{rodrigues2019machine}. How to build an effective MEC system for AI at edge (or edge intelligence) to facilitate tremendous IoT applications and smart city initiatives has become of paramount importance. \textcolor{blue}{Although intensive research targets at task offloading with latency requirements, none of them has really spelled out what tasks to be computed and for what purpose.} Therefore, how to holistically tie both MEC and edge intelligence is of high priority. One of important research directions is how to leverage MEC to boost the federated learning, which is still in its infancy and demands further research \cite{lim2020federated}.

ML typically consumes intensive computing resource, particularly for real-time or delay-sensitive tasks as in connected and autonomous driving, which is why MEC was introduced \cite{rodrigues2019machine}. How to build an effective MEC system for AI at edge (or edge intelligence) to facilitate tremendous IoT applications and smart city initiatives is of paramount importance. Under this design, edge offloading (from MDs to ESs or among ESs) potentially plays a crucial role in providing data input for edge caching, edge training, and edge inference according to the dynamic resources in the MEC systems~\cite{xu2021edge}. By combining MEC with edge intelligence, we could investigate how to leverage MEC to boost the federated learning, which is still in its infancy and demands further research.

\vspace{-0.15in}
\subsection{Develop an Effective Holistic MEC Service Ecosystem}

Although MEC was conceived in response to the edge computing, it turns out MEC as a service system can be leveraged to enable significantly more services beyond computing as we envisioned in \cite{ding2019beef2}. If being built well, MEC can be used to perform sensing and data collection, push delay-tolerant data to the edge to facilitate fast communications, cache popular content of large volume for content distribution, provide temporary buffer/storage for effective transmission scheduling and computing, conserve energy for battery-driven devices, and extract intelligence at edge or at premises where intelligence is needed. This is true for smart city initiatives where omnipresent SCCSI capability is needed for smart city operations, particularly where capability-enabled vehicles, public or private, can be utilized to provide utility-like services \cite{ding2019beef2}. Thus, the future design for MEC should focus on the development of holistic utility-like service ecosystem to enable future intelligentization of life sustaining systems to improve people's quality of life (QoL). This low-cost MEC ecosystem may help ease the digital divide and provide remote education and training services.

\section{Concluding Remarks}
Multi-access edge computing (MEC) provides an effective approach to push computing rich cloud services to the proximity of end users, the so-called edge, in order to facilitate emerging latency-sensitive applications like connected and autonomous driving, smart and connected health, and timely edge intelligence services. Although there exist tremendous research on MEC in the past few years, there are still many design challenges as many interesting applications emerge. In this paper, we attempt to clarify some critical concepts in MEC and realistic design issues, elaborate  the important research issues, and identify future research directions. Our focus is much more on the problem identification and optimization formulation from the technical perspective. It is our high hope that this paper could help readers quickly get familiar with the research critical problems and make significant contributions to solve the important problems in this area.

\section*{Acknowledgements}
The work of X. Chen, G. Zhu and Y. Fang was supported in part by US National Science Foundation under IIS-1722791 and CNS-2106589. The work of X. Deng was supported in part by  National Natural Science Foundation of China (No. 62172441, 61772553).

%TC:ignore
\renewcommand\refname{References}
\bibliographystyle{IEEEtran}
\bibliography{edge}

\begin{thebibliography}{10}
\providecommand{\url}[1]{#1}
\csname url@samestyle\endcsname
\providecommand{\newblock}{\relax}
\providecommand{\bibinfo}[2]{#2}
\providecommand{\BIBentrySTDinterwordspacing}{\spaceskip=0pt\relax}
\providecommand{\BIBentryALTinterwordstretchfactor}{4}
\providecommand{\BIBentryALTinterwordspacing}{\spaceskip=\fontdimen2\font plus
\BIBentryALTinterwordstretchfactor\fontdimen3\font minus
  \fontdimen4\font\relax}
\providecommand{\BIBforeignlanguage}[2]{{%
\expandafter\ifx\csname l@#1\endcsname\relax
\typeout{** WARNING: IEEEtran.bst: No hyphenation pattern has been}%
\typeout{** loaded for the language `#1'. Using the pattern for}%
\typeout{** the default language instead.}%
\else
\language=\csname l@#1\endcsname
\fi
#2}}
\providecommand{\BIBdecl}{\relax}
\BIBdecl

\bibitem{ding2017cognitive2}
H.~Ding, Y.~Fang, X.~Huang, M.~Pan, P.~Li, and S.~Glisic, ``Cognitive capacity
  harvesting networks: Architectural evolution toward future cognitive radio
  networks,'' \emph{IEEE Commun. Surveys Tuts.}, vol.~19, no.~3, pp.
  1902--1923, 2017.

\bibitem{koketsurodrigues2021offloading}
T.~Koketsurodrigues, J.~Liu, and N.~Kato, ``Offloading decision for mobile
  multi-access edge computing in a multi-tiered 6g network,'' \emph{IEEE Trans.
  on Emerg. Topics in Comput.}, to be published, doi:
  10.1109/TETC.2021.3090061.

\bibitem{mao2017survey}
Y.~Mao, C.~You, J.~Zhang, K.~Huang, and K.~B. Letaief, ``A survey on mobile
  edge computing: The communication perspective,'' \emph{IEEE Commun. Surveys
  Tuts.}, vol.~19, no.~4, pp. 2322--2358, Fourthquarter 2017.

\bibitem{abbas2017mobile}
N.~Abbas, Y.~Zhang, A.~Taherkordi, and T.~Skeie, ``Mobile edge computing: A
  survey,'' \emph{IEEE Internet of Things J.}, vol.~5, no.~1, pp. 450--465,
  2017.

\bibitem{mach2017mobile}
P.~Mach and Z.~Becvar, ``Mobile edge computing: A survey on architecture and
  computation offloading,'' \emph{IEEE Commun. Surveys Tuts.}, vol.~19, no.~3,
  pp. 1628--1656, thirdquarter 2017.

\bibitem{yuan2020joint}
Q.~Yuan, J.~Li, H.~Zhou, T.~Lin, G.~Luo, and X.~Shen, ``A joint service
  migration and mobility optimization approach for vehicular edge computing,''
  \emph{IEEE Trans. Veh. Technol.}, vol.~69, no.~8, pp. 9041--9052, Aug. 2020.

\bibitem{han2020joint2}
D.~Han, W.~Chen, and Y.~Fang, ``Joint channel and queue aware scheduling for
  latency sensitive mobile edge computing with power constraints,'' \emph{IEEE
  Trans. on Wireless Commun.}, vol.~19, no.~6, pp. 3938--3951, June 2020.

\bibitem{han2019offloading}
D.~Han, W.~Chen, B.~Bai, and Y.~Fang, ``Offloading optimization and bottleneck
  analysis for mobile cloud computing,'' \emph{IEEE Transactions on
  Communications}, vol.~67, no.~9, pp. 6153--6167, September 2019.

\bibitem{poularakis2020service}
K.~Poularakis, J.~Llorca, A.~M. Tulino, I.~Taylor, and L.~Tassiulas, ``Service
  placement and request routing in mec networks with storage, computation, and
  communication constraints,'' \emph{IEEE/ACM Trans. on Netw.}, vol.~28, no.~3,
  pp. 1047--1060, June 2020.

\bibitem{deng2021throughput2}
Y.~Deng, Z.~Chen, X.~Chen, and Y.~Fang, ``Throughput maximization for
  multi-edge multi-user edge computing systems,'' \emph{IEEE Internet of Things
  J.}, vol.~9, no.~1, pp. 68--79, Jan. 2022.

\bibitem{deng2021leverage}
Y.~Deng, Z.~Chen, X.~Chen, X.~Deng, and Y.~Fang, ``How to leverage mobile
  vehicles to balance the workload in multi-access edge computing systems,''
  \emph{IEEE Trans. on Veh. Technol.}, vol.~70, no.~11, pp. 12\,283--12\,286,
  Dec. 2021.

\bibitem{zhang2021joint}
X.~Zhang, Z.~Li, C.~Lai, and J.~Zhang, ``Joint edge server placement and
  service placement in mobile edge computing,'' \emph{IEEE Internet of Things
  Journal}, 2021.

\bibitem{ding2019beef2}
H.~Ding, Y.~Guo, X.~Li, and Y.~Fang, ``Beef up the edge: spectrum-aware
  placement of edge computing services for the {Internet of Things},''
  \emph{IEEE Trans. Mob. Comput.}, vol.~18, no.~12, pp. 2783--2795, Dec. 2019.

\bibitem{rodrigues2019machine}
T.~K. Rodrigues, K.~Suto, H.~Nishiyama, J.~Liu, and N.~Kato, ``Machine learning
  meets computation and communication control in evolving edge and cloud:
  Challenges and future perspective,'' \emph{IEEE Commun. Surveys Tuts.},
  vol.~22, no.~1, pp. 38--67, firstquarter 2020.

\bibitem{xu2021edge}
D.~Xu, T.~Li, Y.~Li, X.~Su, S.~Tarkoma, T.~Jiang, J.~Crowcroft, and P.~Hui,
  ``Edge intelligence: Empowering intelligence to the edge of network,''
  \emph{Proceedings of the IEEE}, vol. 109, no.~11, pp. 1778--1837, 2021.

\end{thebibliography}

\begin{IEEEbiographynophoto}{Yiqin Deng} received the B.S. degree in project management from Hunan Institute of Engineering, Xiangtan, China, in 2014, and the M.S. degree in
software engineering from Central South University, Changsha, China, in 2017, where she is currently pursuing the Ph.D. degree in computer science and technology. She was a Visiting Researcher with the University of Florida, Gainesville, FL, USA, from 2019 to 2021. Her research interests are primarily focused on edge/fog computing, Internet of Vehicle, and resource management.
\end{IEEEbiographynophoto}

\begin{IEEEbiographynophoto}{Xianhao Chen} received the B.Eng. degree from Southwest Jiaotong University, Chengdu, China, in 2017. Since 2018, he has been working towards the Ph.D. degree with the Department of Electrical and Computer Engineering, University of Florida, Gainesville, FL, USA. His research interests include wireless networking, mobile crowdsourcing, and machine learning.
\end{IEEEbiographynophoto}

\begin{IEEEbiographynophoto}{Guangyu Zhu} received the B.Eng. degree from Xidian University, Xi'an, China, in 2019. Since 2019, he has been pursuing the Ph.D degree with the Department of Electrical and Computer Engineering, University of Florida, Gainesville, FL, USA. His research interests include machine learning, wireless networks, and edge computing.
\end{IEEEbiographynophoto}

\begin{IEEEbiographynophoto}{Yuguang Fang} (F'08) received an MS degree from Qufu Normal University, China in 1987, a PhD degree from Case Western Reserve University in 1994, and a PhD degree from Boston University in 1997. He joined Department of Electrical and Computer Engineering at University of Florida in 2000 as an assistant professor, then was promoted to an associate professor in 2003 and a full professor in 2005 and has been a distinguished professor since 2019.  He holds multiple professorships including the University of Florida Foundation Preeminence Term Professorship (2019-2022), University of Florida Research Foundation Professorship (2017-2020, 2006-2009), and University of Florida Term Professorship (2017-2021). He received the 2018 IEEE Vehicular Technology Outstanding Service Award, the Best Paper Award from IEEE ICNP (2006), and the 2010-2011 UF Doctoral Dissertation Advisor/Mentoring Award. He was the Editor-in-Chief of IEEE Transactions on Vehicular Technology (2013-2017) and IEEE Wireless Communications (2009-2012). He is a fellow of IEEE and AAAS.
\end{IEEEbiographynophoto}

\begin{IEEEbiographynophoto}{Zhigang Chen}
received the B.S., the M.S. and Ph.D. from Central South University in China in 1984, 1987 and 1998, respectively. He is currently a professor with School of Computer Science and Engineering, Central South University. He is also a director and senior member of China Computer Federation (CCF), and a member of pervasive computing committee of CCF. His research interests include cluster computing, parallel and distributed systems, computer security and wireless networks.
\end{IEEEbiographynophoto}

\begin{IEEEbiographynophoto}{Xiaoheng Deng}
received the Ph.D. degrees in computer
science from Central South University, Changsha,
Hunan, P.R. China, in 2005. Since 2006, he has
been an Associate Professor and then a Full Professor
with the Department of Electrical and Communication
Engineering, Central South University.He
is Joint researcher of Shenzhen Research Institue,
Central South University. He is a senior member
of CCF, a member of CCF Pervasive Computing
Council, a member of IEEE and ACM. He has been
a chair of CCF YOCSEF CHANGSHA from 2009
to 2010. His research interests include edge computing, Internet of Things,
online social network analysis, data mining, and pattern recognization.
\end{IEEEbiographynophoto}

%TC:endignore
\end{document}